\documentclass[aps,prl,twocolumn,showpacs,superscriptaddress]{revtex4-2}
\usepackage{graphicx}
\usepackage{dcolumn}
\usepackage{bm}
\usepackage{hyperref}
\usepackage[T1]{fontenc}
\usepackage[utf8]{inputenc}
\usepackage[english]{babel}
\usepackage{units}

\usepackage{mathtools}
\hypersetup{colorlinks=true,citecolor={blue},linkcolor={blue},urlcolor={blue}}
\usepackage{amsmath}
\usepackage{empheq}
\usepackage{mathrsfs}
\usepackage[normalem]{ulem}
\usepackage{amssymb}
\usepackage{soul}
\usepackage{upgreek}
\usepackage{xcolor}

\begin{document}

\title{Tailoring the resonant spin response of a stirred polariton condensate}

\author{I. Gnusov}
\email[]{Ivan.Gnusov@skoltech.ru}
\address{Skolkovo Institute of Science and Technology, Moscow, Territory of innovation center “Skolkovo”,
Bolshoy Boulevard 30, bld. 1, 121205, Russia.}

\author{A.Yulin}
\address{Department of Physics, ITMO University, Saint Petersburg 197101, Russia}

\author{S. Baryshev}
\address{Skolkovo Institute of Science and Technology, Moscow, Territory of innovation center “Skolkovo”,
Bolshoy Boulevard 30, bld. 1, 121205, Russia.}

\author{S. Alyatkin}
\address{Skolkovo Institute of Science and Technology, Moscow, Territory of innovation center “Skolkovo”,
Bolshoy Boulevard 30, bld. 1, 121205, Russia.}

\author{P. G. Lagoudakis}
\address{Skolkovo Institute of Science and Technology, Moscow, Territory of innovation center “Skolkovo”,
Bolshoy Boulevard 30, bld. 1, 121205, Russia.}

\begin{abstract}
We report on the enhancement of the spin coherence time ($T_2$) by almost an order-of-magnitude in exciton-polariton condensates through driven spin precession resonance. Using a rotating optical trap formed by a bichromatic laser excitation, we synchronize the trap’s stirring frequency with the condensate’s intrinsic Larmor precession, achieving an order of magnitude increase in spin coherence. By tuning the optical trap profile via excitation lasers intensity, we precisely control the resonance width. Here we present a theoretical model, that explains our experimental findings in terms of the mutual synchronization of the condensate circular polarization components. Our findings underpin the potential of polariton condensates for spinoptronic devices and quantum technologies.


\end{abstract}

\maketitle 

The nuclear magnetic resonance (NMR) effect~\cite{nmr_rabi} finds numerous applications in medicine, material science, and quantum computing. NMR builds upon the spin Larmor precession wherein the spin put in the magnetic field precesses around it. The resonant spin response appears when the radio frequency (RF) field applied to the studied system is exactly in resonance with the Larmor precession frequency. The NMR effect is a handy tool for quantum computing applications, where the RF field of controllable frequency and duration is used for the preparation of the required spin state. In this regard, the physical systems with high spin coherence time ($T_2$) are prominent for applications requiring multiple qubit operations. Recently, an effect analogous to conventional NMR was discovered for exciton-polariton (polaritons further on) condensates~\cite{gnusov2024observation}.

Polaritons are composite quasiparticles formed in semiconductor microcavity by the strong coupling of photons and excitons~\cite{kavokin_microcavities_2007}. Due to their hybrid nature, polaritons inherit remarkable properties such as low effective mass, mutual nonlinear interactions, superfluidity and ability to form Bose condensates even at room temperature~\cite{Zasedatelev2019}. Angular momentum conservation couples cavity photons and excitons with the same spin projection ($\pm\hbar$) onto the cavity growth axis. The resultant polariton state is characterized by an integer pseudospin (or simply "spin") corresponding explicitly to circular polarization of the emitted cavity light. The large Coulomb interaction inherent to polaritons makes them prominent for spinoptronic  devices~\cite{Amo_NatPho2010, Cerna_NatComm2013, askitopoulos_all-optical_2018} which combine the best features of light and spin-sensitive materials~\cite{pol_review, Liew_PhysE2011}. Until now, numerous nonlinear polariton spin effects have been reported including bistability~\cite{pickup_optical_2018, Ohadi19, Sigurdsson_PRR2020}, bifurcations~\cite{ohadi_spontaneous_2015}, optical spin Hall effect~\cite{leyder_observation_2007}, half-quantised vortices~\cite{Lagoudakis2009_halfquantised}, and the self-induced Larmor precession~\cite{shelykh_spin_dyn_2004,larmor_polariton2006, baryshev_prl, coher_revivals}. The distinct advantage of polaritons is the ability to utilize only optical fields for controlling the spin state. For instance, the recently reported effect of optically driven spin precession~\cite{gnusov2024observation} offers a new avenue for the addressable control of polariton spin analogously to the NMR effect. However, the observed spin coherence time $T_2$ of 320 ps could be the factor that limits future applications.

In this Letter, we investigate the driven spin precession resonance of polariton condensate and propose an approach to controllably increase its spin coherence. For this, we realize the rotating optical trap which stirs the polariton condensate on a par with its spin. At GHz stirring frequencies, when the external stirring matches the intrinsic condensate Larmor precession frequency, the condensate demonstrates an increased spin coherence. We show that the width of the precession resonance depends on the shape of the rotating potential, defined by the variable composition ratio of the two excitation laser beams. Moreover, we report an order-of-magnitude increase of the spin coherence ($T_2$) time with a sharp precession resonance observed at extremely small (0.5$\%$) time-periodic modulation of the confining potential. To explain the experimental findings, we develop a theory allowing efficient description of the polariton dynamics at the pumps of intensities well above the condensation threshold.

\begin{figure}[hbt]
    \centering
    \includegraphics[width=0.99\columnwidth]{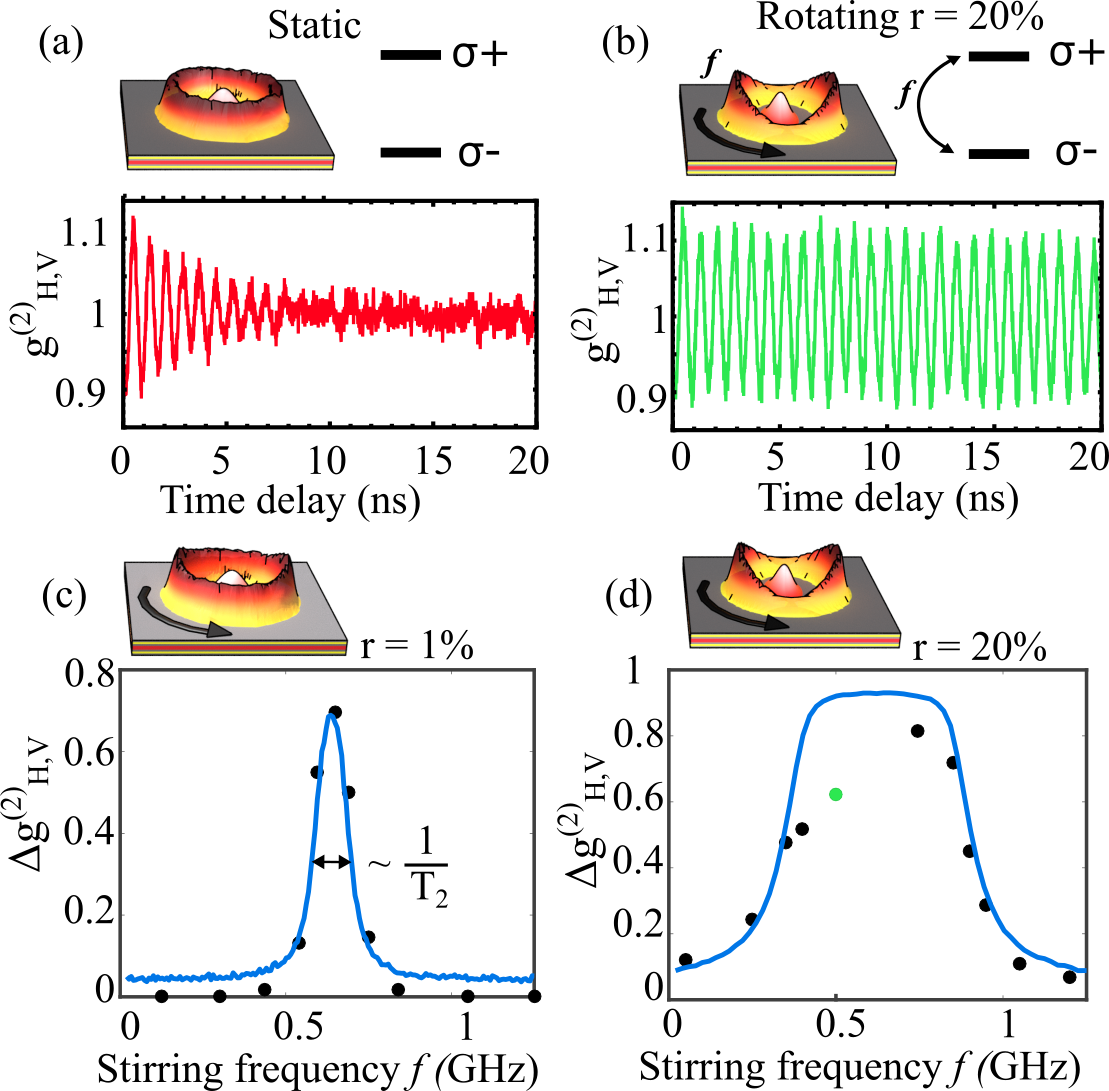}
    \caption{(a) Measured $g^{(2)}_{H,V}$ for the polariton condensate in a static annular optical trap at pump power $P= 2.7$ $P_{th}$. (b) Measured $g^{(2)}_{H,V}$ for the condensate in the rotating optical trap ($ r = 20 \%$) at $f=0.5$ GHz resonant to the self-induced Larmor precision frequency. The upper inset in panel (b) showcases the shape of the optical pump and induced energy splitting between the cross-polarized spin components of the condensate that is equal to the frequency of the external time-periodic drive. (c) The amplitude of the spin precession $\Delta g^{(2)}_{H,V}$ in the vicinity of the 15 ns time delay for (c) $ r = 1 \%$ and (d) $ r =20 \%$ demonstrating the resonant behavior. The upper insets in panels (c,d) schematically represent the shape of the rotating optical pattern for different $r$. The green point in (d) is an amplitude of the spin precession presented in (b).}
    \label{fig1}
\end{figure}

For our experiments, we use an inorganic $2 \lambda$ GaAs/AlAs$_{0.98}$P$_{0.02}$ microcavity with embedded InGaAs quantum wells~\cite{cilibrizzi_polariton_2014} held at 4K. Polariton condensate is excited with the rotating bichromatic beam, created with two single-mode, frequency-detuned and stabilized lasers~\cite{rotatingbucket,fraser2023}. Each of the laser beams is shaped with a reflective spatial light modulator (SLM) into a ring with the phase winding~\cite{Chen2013_perfectvortex} (orbital angular momenta $l_{1,2}=\pm1$ for the first and second laser, respectively). The spatially modulated laser emission is then superimposed with a nonpolarizing beam splitter. The resultant optical pattern presented in Fig.~\ref{fig1}(b) rotates at frequency $f= (f_1-f_2)/(l_1-l_2)$, where  $f_{1,2}$ are the frequencies of the excitation lasers, the sign of $f$ defines the stirring direction. The intensities of the lasers composing the bichromatic beam determine the excitation pattern for polaritons. We fix the intensity of the first laser at $I_1=2.7$$P_{th}$, where $P_{th}$ is condensation threshold power, retrieved for the static (non-rotating) annular optical trap, schematically depicted in  Fig.~\ref{fig1}(a). The admixture of the second laser with the intensity $I_2$ alters the shape of the excitation pattern from more to less uniform around the circumference (compare the insets in Fig.~\ref{fig1}(c) and Fig.~\ref{fig1}(b)). Below, we focus on the driven spin precession of the condensate created in traps of different shape, controlled by the intensity ratio $r = I_2 / I_1$ varied in the range from 0.5 $\%$ to 30 $\%$. Here, the biggest $r$ corresponds to the dumbbell excitation potential, as shown in Fig.~\ref{fig1}(b), and the smallest $r$ results in an almost uniform ring-like potential with the small time-periodic intensity modulation on top.

The spin of the condensate is encoded in the polarization of the photons created when polaritons decay~\cite{kavokin_microcavities_2007}. To study the spin of the condensate we split polariton photoluminescence (PL) with the polarizing beamsplitter (PBS) in the linear polarization basis and employ the Hanbury Brown and Twiss (HBT) interferometry~\cite{baryshev_prl}. Namely, we investigate the cross-correlation function $g^{(2)}_{H,V}$ between the horizontal $I_H$ and vertical $I_V$ polarization projections of the condensate emission defined as 

\begin{align}\label{eq.g2}
g_{H,V}^{(2)}(\tau)= \frac{\langle I_H(t) I_V(t+\tau)\rangle}{\langle{I_H(t)}\rangle\langle{I_V(t)}\rangle}. 
\end{align}

The cross-correlation is measured utilizing the time-correlated single photon counting technique (TCSPC) and avalanche photodiodes as detectors. As a result, we obtain the cross-correlation $g^{(2)}_{H,V} (\tau)$ as a function of the time delay $\tau$ between the HBT interferometer arms that allows us to track the recurrent spin dynamics in short time scales.

 Excitation of the condensate with the elliptically polarized laser gives rise to the imbalance of the opposite spin polariton population.  The polariton-polariton and polariton-reservoir interactions result in the different blueshifts of cross-polarized condensates leading to the energy splitting, as shown schematically in Fig.~\ref{fig1}(a). The emergent splitting causes the self-induced Larmor precession~\cite{larmor_polariton2006,Askitopoulos_robust} of the condensate spin. The measured cross-correlation $g^{(2)}_{H,V} (\tau)$ presented in Fig.~\ref{fig1}(a) clearly shows such a precession with the oscillations rapidly decaying in time. The envelope of the oscillations characterises the stability of the precession detrimentally affected by the polariton interactions with the incoherent reservoir. In the case of the static trap the decay time of the oscillations (spin dephasing time) can be as long as 5 ns~\cite{baryshev_prl}. 

On the contrary, we find that the precession becomes long-lasting when the trap is stirred externally, as visible in Fig.~\ref{fig1}(b). For the intensities ratio of $ r =20 \%$, the rotating excitation pattern is cylindrically asymmetric (see the inset in Fig.~\ref{fig1}(b)), and therefore induces an effective elliptical confinement and a torque for the condensate. The interplay of the cavity inherent TE-TM splitting and trap spatial ellipticity makes the condensate adopt the defined linear polarization (spin) aligned to the short axis of the confinement potential~\cite{gnusov_prapl}. Furthermore, when the frequency of the stirring is equal or close to the condensate self-induced precession frequency the spin starts to rotate in step with the optical trap with the renewed stability (see green curve in Fig.~\ref{fig1}(b)). In this case, the spin dephasing time is limited by the mutual frequency stability of the excitation lasers (a few $\mu$s in our experiments) rather than by the fluctuations in the polariton system. Analogously to the conventional NMR, this effect can be described by the effective magnetic field coming from the trap rotation and acting on the condensate spin~\cite{gnusov2024observation}. 

Our experiments reveal that the shape of the rotating excitation pattern, governed by the intensity ratio $r$ of two excitation lasers strongly affects the width of the observed precession resonance. Figure~\ref{fig1}(c,d) depicts the measured resonant response of our system to the external stirring $f$ for the rotating traps with $r=1\%$ and $r = 20\%$, respectively. Namely, the curves demonstrate the range (amplitude) of the cross-correlation function $\Delta g^{(2)}_{H,V}$ at a big (15 ns) time delay corrected on the frequency bandwidth of the experimental setup (see Supplementary Materials). For both cases, the resonance is observed near $f = 0.6$ GHz, which is the frequency of the self-induced Larmor precession. Notably, the resonance for $r  = 20\%$ is much wider than that of  $r  = 1\%$ (FWHM~$\approx$ 0.5 GHz and 0.1 GHz, respectively). Let us mention that in this Letter we use the term "resonant curve" for the dependencies of $\Delta g^{(2)}$ on $f$ in the vicinity of the frequency which provides resonant coupling between the $H$ and $V$ polarizations. Therefore, the curves in Fig.~\ref{fig1}(c,d) are not Lorentzians.

\begin{figure*}[t]
    \centering
    \includegraphics[width=0.99\textwidth]{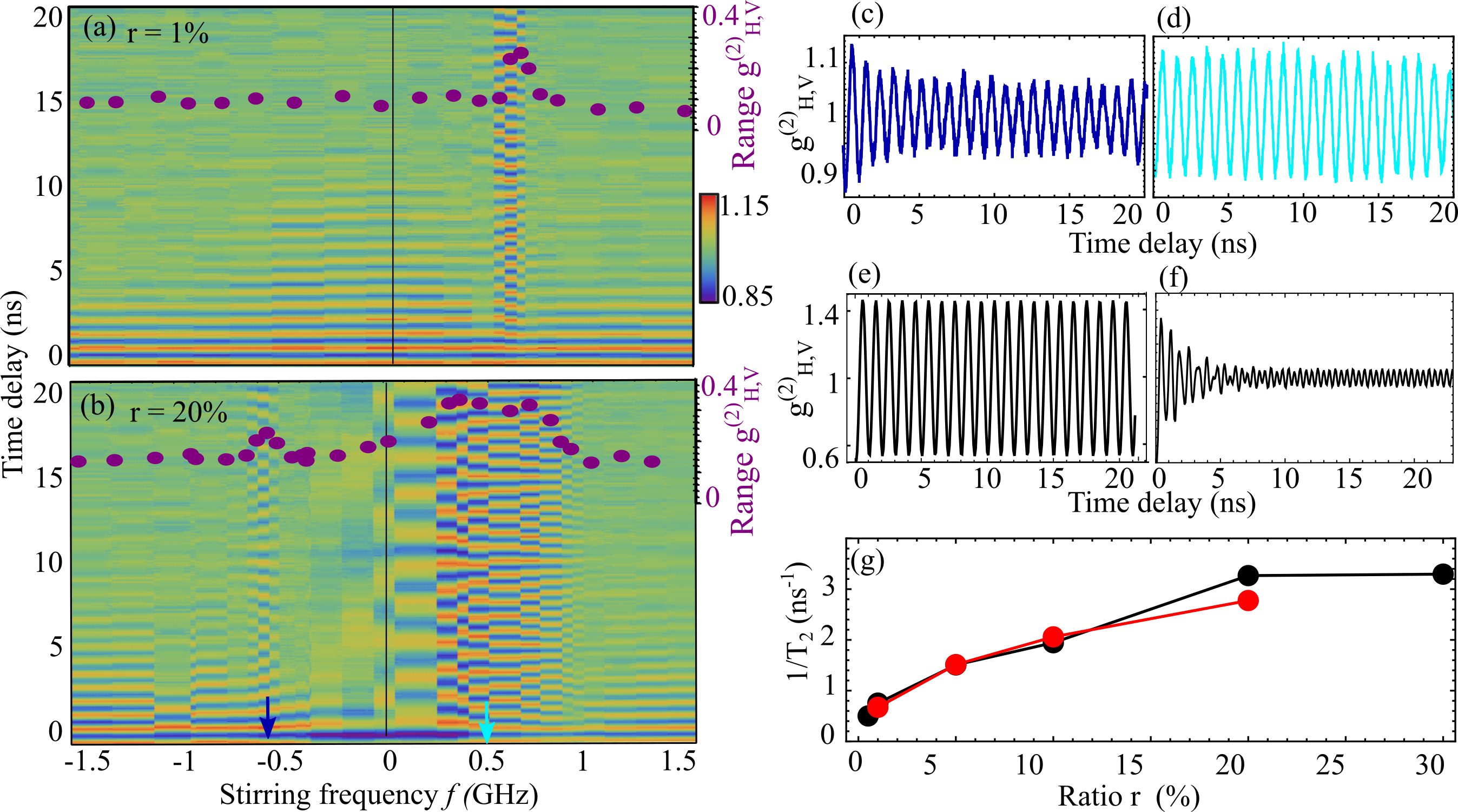}
    \caption{$g^{(2)}_{H,V}$ as a function of the stirring frequency $f$ and time delay of the HBT interferometer for (a) $r= 1\%$ and (b) $r= 20\%$. The purple dots in panels (a,b) represent the range of the $g^{(2)}_{H,V}$ retrieved within 2 ns vicinity of 15 ns time delay. The cross sections of panel (b) taken at $f = -0.5 $ GHz and $f = 0.5 $ GHz are depicted in (c) and (d), respectively. The light and dark blue arrows in (b) mark the positions of the cross-sections. Simulated auto-correlation functions $g^{(2)}_{HV}$ for (d) $f=0.5$ GHz and (f)  $f=1.1$ GHz, the inter-polarization coupling strength is $\tilde\sigma=542$ MHz. (g) The experimental width ($1/\ T_2$) of the resonant curve as a function of $r$. The black dots are experimentally measured widths, the red dots - the widths of the synchronization ranges calculated from (\ref{phase_equation}). The black and red lines are to guide the eye. }
    \label{fig2}
\end{figure*}

The resonant curves in Figs.~\ref{fig1}(c) and \ref{fig1}(d) are retrieved from the measured $g^{(2)}_{H,V}$ cross-correlation maps depicted in Figs.~\ref{fig2}(a) and \ref{fig2}(b). These maps represent the $g^{(2)}_{H,V}$ of the stirred condensate as a function of time delay and stirring frequency $f$ for $r = 1 \%$ and $20\%$, respectively. The spin precession resonance is evident as a $g^{(2)}_{H,V}$ oscillations persistent for the whole range of measured time delays (20 ns). The time series of the spin oscillations for $r = 20 \%$ at $f = \pm 0.5$ GHz are presented in Figs.~\ref{fig2}(c,d). For the the small $r$ the resonance is observed only for the positive stirring direction ($f > 0 $), whereas for the negative stirring frequency ($f < 0 $) the condensate does not adopt the external rotation and only the self-induced Larmor precession with the characteristic precession decay of 5 ns is observed. The direction of the external stirring should correspond to that of the self-induced Larmor precession, that is in turn defined by the sign of the excitation polarization ellipticity (the splitting between cross-circularly polarized states). On the other hand, for the $ r = 20\% $ the small induced precession revival is observed for the negative stirring direction. While the major time-periodic modulation of the excitation potential occurs in the counter-propagating direction, the non-uniformity of the two excitation beams results in a reminiscent stirring co-directed with the self-induced spin precession, causing such, relatively weak, precession revival. This is additionally confirmed by the experimental results obtained for the higher $r = 30\%$, where the resonance is even wider and spans for both negative and positive stirring directions (see Supplementary Materials).

Further, in the experiment we investigate the dependence of the precession resonance width (or the width of the synchronization range in inverse $T_2$) on the shape of the rotating optical trap by varying the intensity ratio $r$ of the excitation lasers from 0.5 to 30 $\% $. We characterise the precession resonance analogously to NMR by $T_2 \approx (FWHM_f\times \pi)^{-1}$ where $FWHM_f$ is the full width at half maximum of the resonance curve~\cite{spin_book2001,kittel}. Thus, $T_2$ is the characteristic time of the transverse spin relaxation that quantifies the "sensitivity" of the condensate spin to external stirring. The experimental results are presented in Fig.~\ref{fig2}(g). We observe a remarkable increase of $T_2$ reaching up to 2 ns for $r = 0.5\%$. So by altering the shape of the stirring potential, we can directly access and improve the coherence of the induced spin precession. Moreover, even a slight admixture of the stirring ($r = 0.5\%$) on top of the static confining potential makes the condensate adopt the external drive. This opens a great perspective for the precise and low-energy control of the condensate spin for the spinoptronics applications.

To explain the variation of the precession resonance width with respect to the shape of the rotating trap, we adopt a mathematical model introduced in Refs. \cite{gnusov2020optical,gnusov2024observation} characterizing the polarization in terms of their slowly varying complex amplitudes $A_{\uparrow,\downarrow}$, where $\uparrow$ denotes clock- and $\downarrow$ - counter-clockwise circular polarizations. 

Assuming the nonlinear effects to be small compared to linear interactions the problem can be reduced to quasi-linear dynamics of the supermodes (the eigenstates found in the presence of linear cross-polarization interaction) \cite{rotatingbucket,yulin2023vorticity,yulin2024persistent,gnusov2024observation}. However, for a relatively high pump ($P=2.7P_{th}$), the nonlinear effects become strong and can dominate over the coupling between the polarizations. Then the quasi-linear approach mentioned above becomes invalid and another perturbation method is required. 
For this, we seek stationary states neglecting the linear cross-polarization interactions. Then the densities of the polaritons in both polarizations are predominantly determined by the pump and the linear interactions can cause only small variations. In this case a well-known Kuramoto approximation can be developed reducing the dynamics to the one of mutual phase between the polarizations.

In our experiments the ellipticity of the pump is small and so the intensity of the polarizations are similar $|A_{\uparrow}| \approx |A_{\downarrow}|=\rho_0$. Then we seek the approximate solution in the form $A_{\uparrow, \downarrow}=\rho_{0}e^{i(\omega_0 \pm \Omega) t +i\phi_{\uparrow,\downarrow}} $ where $\omega_0=\frac{\omega_{\uparrow}+\omega_{\downarrow}}{2}$, $\omega_{\uparrow,\downarrow}$ are the free-running frequencies of non-interacting polarizations, $\Omega$ is the angular velocity of the rotating pump, $\phi_{\uparrow,\downarrow}$ are the phases of the polarizations. In Supplementary Materials we show that an Adler equation can be derived for the mutual phase $\varphi=\phi_{\uparrow}-\phi_{\downarrow}$ 
\begin{eqnarray}
    \dot \varphi=\delta + \tilde \sigma \sin(\varphi) 
    \label{phase_equation}
\end{eqnarray}
where $\delta=\omega_{\uparrow}-\omega_{\downarrow}  -2\Omega$ is the detuning and $\tilde \sigma$ is the effective coupling. Let us note that because of the noise in the pumps (the lasers intensity fluctuations) the detuning $\delta$ is a fluctuating parameter.

Equation (\ref{phase_equation}) has the parameter $\tilde \sigma$ that has to be fitted by comparison with the experimental data. The relation of $\tilde \sigma$ to the parameters of coupled modes equations is discussed in Supplementary Materials in detail.

The equation (\ref{phase_equation})  can be easily studied numerically and analytically. In fact, the evolution of the phase $\varphi$ is described by the equation identical to the one describing the viscous motion of a particle in a wash-board potential, see additional discussion in Supplementary Materials. Moreover, this equation contains all the information necessary to describe the rotation of the polarization plane.

In the leading approximation order, the PL intensity measured through a linear polarizer (analogously to the experimental measurement technique) angled by $\chi$ (where $\chi=0$ and $\chi=\frac{\pi}{2}$ are horizontally and vertically oriented polarizers) is written as  

$I_{\chi}=|A_{\uparrow}e^{-i\chi}+A_{\downarrow}e^{i\chi}|^2 =2\rho_0^2 J(t, \chi)$
where 
\begin{eqnarray}
J= (1+\cos(2\Omega t +\varphi-2\chi)).
\label{J}
\end{eqnarray}

Then for the  $g^2$ correlation function we obtain
\begin{eqnarray}
 g^{(2)}(\tau, \chi_1, \chi_2)= \frac{< J(t, \chi_1)J(t-\tau, \chi_2)> }{< J(t, \chi_1) ><  J(t-\tau, \chi_2) >}.
\label{g2_through_phase}
\end{eqnarray}
Then, the expressions (\ref{J}) and (\ref{g2_through_phase}) allow to calculate the correlation function from the known temporal evolution of the mutual phase $\varphi$. The correlation functions $g^{(2)}_H$ and $g^{(2)}_{HV}$ are defined as $g^{(2)}_H = g^{(2)}(\tau, 0, 0)$ and $g^{(2)}_{HV}=g^{(2)}(\tau, 0, \pi/2)$. 

\begin{figure}[hbt]
    \centering
    \includegraphics[width=0.99\columnwidth]{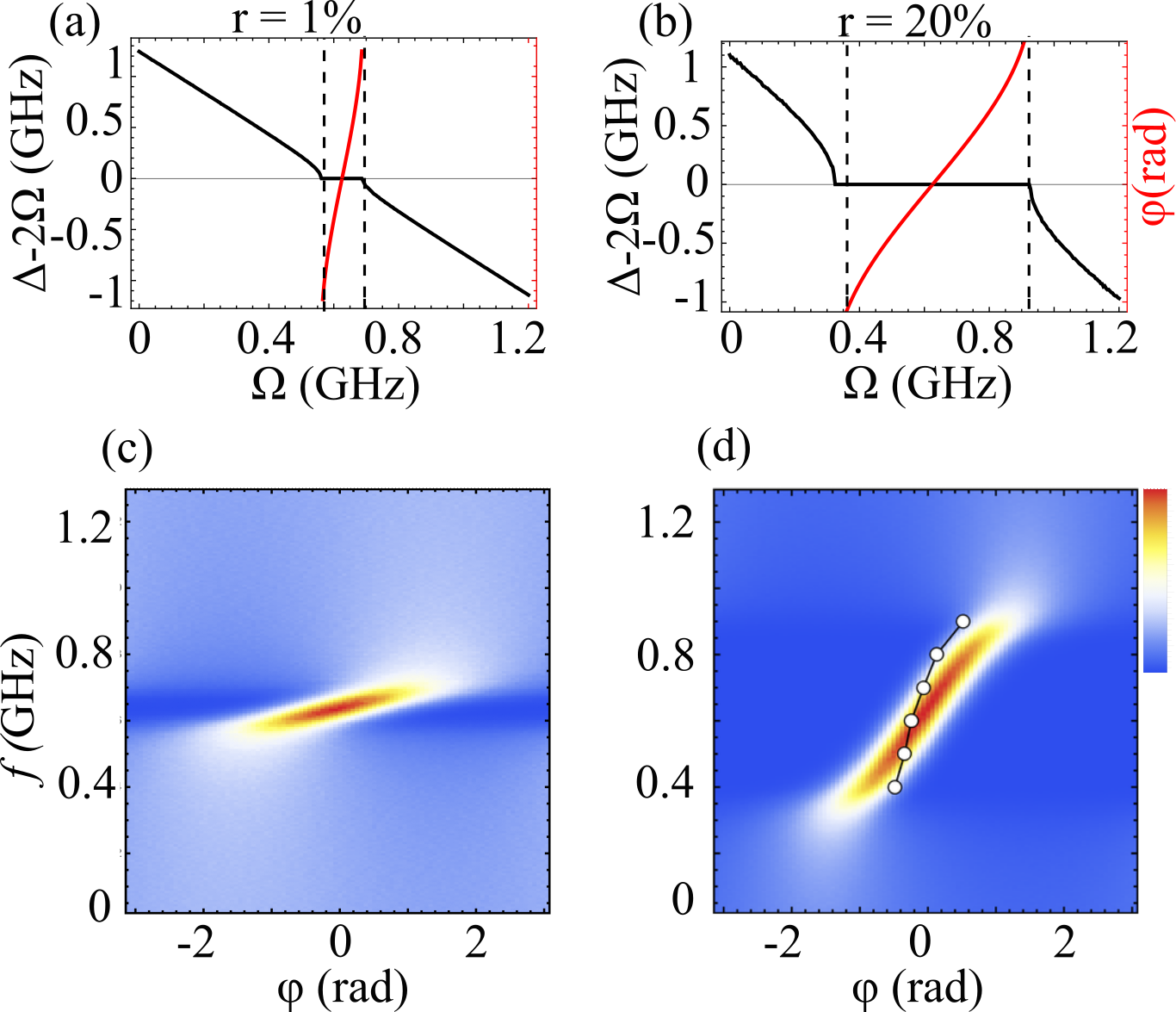}
    \caption{(a,b) The dependence of the average detuning $\Delta=\omega_{\uparrow}-\omega_{\downarrow} + < \dot \varphi > $ of the frequencies of $\uparrow$ and $\downarrow$ polarizations from the potential rotation velocity $\Omega$ for the ratio $r= 1$\% ($\tilde\sigma=144$ MHz) and $20$\% ($\tilde\sigma=542$ MHz), respectively. The synchronization range in (a,b) is shown with vertical dashed lines. The red lines in (a,b) depict stationary mutual phase $\varphi$ as a function of $\Omega$ within the synchronization range. Plots in panels (a,b) are obtained from the numerical simulations of Eq.~\ref{phase_equation}.  (c,d) The numerically calculated probability density of the mutual phase $\varphi$ as a function of $\Omega$ for the $r =20$ \% ($\tilde\sigma=542$ MHz)  and $1$\% ($\tilde\sigma=144$ MHz), respectively. The experimentally measured mutual phases are shown with the white circles in panel (d).}
    \label{Suppl}
\end{figure}

In the following, we first discuss the dynamics of $\varphi$ in the absence of the noise. For large detuning $|\delta| \gg |\tilde \sigma|$ the approximate solution is $\varphi \approx \delta t +\varphi_0$, where $\varphi_0$ is a constant. The linear growth of the mutual phase $<\dot \varphi>=\delta$ which means that the frequency difference between the polarizations is equal to the difference of their free-running frequencies $\omega_{\uparrow}-\omega_{\downarrow}$. However, in the range $|\delta|<|\tilde \sigma|$ the polarizations are locked with $\varphi=\text{arcsin}( \frac{\delta}{\tilde \sigma} )$ and their frequency difference is exactly equal to $2\Omega$. The dependencies of the detuning of the pseudo-spin precession frequency from $2\Omega$ on $\Omega$ are shown in Figs.~\ref{Suppl}(a,b) for different coupling strengths ($r = 1\%$ and $r=20\%$, respectively). Within the synchronization range, the difference is equal exactly to zero. The stationary mutual phase $\varphi$ does not depend on time and is a function of $\Omega$, see the red curves in Figs.~\ref{Suppl}(a,b). Moreover, these results are also well reproduced by the modelling of the Adler equation (\ref{phase_equation}) with the experimental parameters.  

Then, we model the equation (\ref{phase_equation}) with fluctuating detuning $\delta$ and calculate the probability densities for different rotation frequencies $\Omega = 2\pi f$. We find that within the synchronization range the phase fluctuates around the equilibrium value (see Figs.~\ref{Suppl}(c) and \ref{Suppl}(d)). The width of the probability distribution depends on the intensity of the noise. The probability function gets wider closer to the edges of the synchronization domain and is nearly evenly distributed outside the synchronization range. 

We note that the phase $\varphi$ defines the angle between the symmetry axis of the pump and the direction of the polarization plane of the polaritons and thus can be also retrieved experimentally. The experimentally measured phases between the circular polarizations are shown in Fig.~\ref{Suppl}(d), and the developed theory reproduces the experimental observations.

The noise results in the relaxation of the correlation function to the slowly decaying plateau. This is confirmed by the $g^{(2)}_{HV}$ versus time delay obtained in both experiment and numerical simulation (see panels (c)-(f) in Fig.~\ref{fig2}). For the experimental parameters the correlation function decays at times exceeding a hundred of nanoseconds, and the time was limited by the frequency stability of the excitation lasers as reported in \cite{gnusov2024observation}.   

With the developed theory we also reproduce the resonance curves presented in Figs.~\ref{fig1}(c,d). Notably, the amplitude of spin oscillations is large in the range of $\Omega$, where synchronization occurs, which underlines the high degree of condensate polarization and the efficiency of the stirring induced spin rotation. We note that the developed theory also allows to find the spectra of the signal measured through a linear polarizer and links the width of the synchronization range to the width of the spectra (see Supplementary Materials for details). Moreover, the simulations quantitatively reproduce the change of the spin coherence time ($T_2$) depending on the shape of the stirring potential in  Fig.~\ref{fig2}(g) and resonant curves presented in Figs.~\ref{fig1}(c) and \ref{fig1}(d).

To sum up, we have investigated both numerically and experimentally the driven spin precession resonance of the stirred polariton condensate in optically induced trap. By varying the intensity ratio  $r$ of the excitation lasers we change the shape of the stirring potential. The bigger modulation results in a wider precession resonance. Remarkably, the driven spin precession is present even for the very small modulation of the optical trap intensity ($r= 0.5 \%$) allowing for the efficient control of the condensate spin. We have developed a theoretical model, that reproduces our experimental findings and explains them in terms of the mutual synchronization of the circular polarization components. 
 
The almost an order increased spin coherence ($T_2$) of the polariton condensates offers prospects for their use in quantum computing~\cite{Kavokin2022,Barrat2024} and realization of polaritons spin qubits ~\cite{spinqub}. Moreover, analogously to NMR, our findings will potentially allow for the addressable control of individual condensate spin  in a chain or in a lattice~\cite{PhysRevLett.119.067401}, and could enable locking or controlling the state of the array of interacting condensates by resonantly stirring the spin of targeted condensates. Moreover, the rotation of the potential locks the polarization of the condensate and greatly increases the coherence time of the pseudo-spin precession. Thus, the considered polariton system looks promising for studying time crystals~\cite{Zhang2017}. Future investigations of these objects will allow to delve into Floquet physics realized with polaritons~\cite{delValleInclanRedondo2024}.

\section{Acknowledgments}
This work was supported by Russian Science Foundation (RSF) grant no. 24-72-10118.

\setcounter{equation}{0}
\setcounter{figure}{0}
\setcounter{section}{0}
\renewcommand{\theequation}{S\arabic{equation}}
\renewcommand{\thefigure}{S\arabic{figure}}
\renewcommand{\thesection}{S\arabic{section}}
\onecolumngrid
\newpage
\vspace{1cm}

\begin{center}
\Large \textbf{Supplementary Materials: Tailoring the resonant spin response of a stirred polariton condensate}
\end{center}

\section{S1. The derivation of the Adler equation for mutual phase between the circular polarizations of the condensate}

The dynamics of the condensate forming under the action of an incoherent pump can be described in terms of the amplitudes of the fundamental modes of the clock- and counter-clockwise polarizations. This approach is based on the assumption that the effective potential acting on the condensate consists of a main part trapping the condensate and a small correction. Then, the condensate can be represented as an expansion over the modes of the main part of the potential. The small correction to the potential is then accounted for by a slow variation of the amplitudes of the eigenmodes of the unperturbed problem. 

Different modes can have different dissipation rates that can be negative. The latter case corresponds to the condensation. The condensation can occur in several modes simultaneously resulting in the formation of the condensates with several fractions having different frequencies. These multi-frequency regimes are out of the scope of this work, where we focus on the case where the competition between the modes is such that only the ground state survives. However, the other modes are not exactly equal to zero because the spatially non-uniform perturbation of the potential introduces the interaction between the modes and thus the higher modes are excited non-resonantly by the scattering from the ground states. If the detunings of the eigenfrequencies of the modes are large enough, then the non-resonant modes can be adiabatically excluded but then the coupling between the ground states appears. We consider this case and for the sake of simplicity analyze the case of fast relaxation of the reservoir.

In the presence of an angular-dependent potential $\sim \cos(  2 \Theta) $, where $\Theta$ is the angular coordinate, the equations for the slowly varying amplitudes $A_{\uparrow,\downarrow}$ of the ground states in the clock ($\uparrow$ polarization) and counter-clockwise polarization ($\downarrow$ polarization) can be written as

 \begin{eqnarray}
    \dot A_{ \uparrow, \, \downarrow}=\left( -\gamma  +(1+i\alpha) { \cal R_{ \uparrow, \, \downarrow} } +i \tilde \alpha { \cal R_{ \downarrow, \, \uparrow} }  +i\mu |A_{ \uparrow, \, \downarrow}|^2 +i\tilde \mu |A_{ \downarrow, \, \uparrow}|^2 \right) A_{ \uparrow, \, \downarrow} + (1+i \alpha)\frac{\sigma}{\sqrt{1+\bar \alpha^2}} e^{\pm 2i\Omega t}A_{\downarrow, \, \uparrow}, 
     \label{main_equation_S} 
\end{eqnarray}

where $\gamma$ accounts for the linear losses and ${\cal  R_{\uparrow, \downarrow} }=\frac{P_{ \uparrow, \, \downarrow}}{1+|A_{ \uparrow, \, \downarrow}|^2}$ are the densities of the incoherent exciton reservoirs produced by the pumps $P_{\uparrow,\downarrow}$. The reservoirs $R_{\uparrow,\downarrow}$ produce the gain for the polaritons with the polarization coinciding with the polarization of the excitons. The ratios of the blueshift to the gain produced by the excitons are $\alpha$ (for the polaritons with the polarization of the excitons) and $\tilde \alpha$ (for the polaritons of opposite polarization). The nonlinear interaction between the polaritons of the same polarization shifts their frequencies by the value proportional to their density multiplied by the coefficient $\mu$. The coefficient $ \tilde \mu$ accounts for the blueshift experienced by the polaritons of one polarization due to nonlinear interactions with polaritons of the opposite polarization. The composite effect of TE-TM splitting and the presence of the angle-dependent potential couples the ground state with the interaction strength $\sigma$. The coupling is complex and by a simple perturbation theory it can be shown that for the chosen shape of the potential this ratio is  equal to $\alpha$ in the leading approximation order.  The angle-dependent potential is rotating with the angular velocity $\Omega$, thus the coupling is time-dependent.

It is convenient to write equations (\ref{main_equation_S}) in terms of the real amplitudes $\rho_{ \uparrow, \, \downarrow} $ and the phases $\theta_{ \uparrow, \, \downarrow}$ of the fields; $A_{ \uparrow, \, \downarrow}=\rho_{ \uparrow, \, \downarrow} \exp( \pm i \Omega t +i\theta_{ \uparrow, \, \downarrow})$. The equations read

 \begin{eqnarray}
    \dot \rho_{ \uparrow, \, \downarrow}=-\gamma \rho_{ \uparrow, \, \downarrow} +\frac{P_{ \uparrow, \, \downarrow}}{1+\rho_{ \uparrow, \, \downarrow}^2}\rho_{ \uparrow, \, \downarrow}  +\frac{\sigma}{\sqrt{1+ \alpha^2}} \rho_{ \downarrow, \, \uparrow} \left( \cos(\theta_{ \downarrow, \, \uparrow} - \theta_{ \uparrow, \, \downarrow} )- \alpha  \sin(\theta_{ \downarrow, \, \uparrow} - \theta_{ \uparrow, \, \downarrow})  \right)      \label{ampl_phase_eq_S_ampl} \\
    \dot \theta_{ \uparrow, \, \downarrow}=\mp \Omega  + \alpha \frac{ P_{ \uparrow, \, \downarrow} }{1+\rho_{ \uparrow, \, \downarrow}^2}+\mu \rho_{ \uparrow, \, \downarrow}^2 + \frac{\tilde \alpha P_{ \downarrow , \, \uparrow}}{1+\rho_{\downarrow , \, \uparrow}^2} +\tilde \mu \rho_{\downarrow , \, \uparrow}^2 + \frac{\sigma \rho_{\downarrow , \, \uparrow}}{\rho_{ \uparrow, \, \downarrow}\sqrt{1+ \alpha^2}} \left(  \sin(\theta_{ \downarrow, \, \uparrow} - \theta_{ \uparrow, \, \downarrow}) +  \alpha \cos(\theta_{ \downarrow, \, \uparrow} - \theta_{ \uparrow, \, \downarrow}) \right).
    \label{ampl_phase_eq_S_phase}
\end{eqnarray}

Without the interaction ($\sigma=0$) the stationary real amplitudes can easily be found $\rho_{0 \uparrow, \, \downarrow}=\sqrt{\frac{P_{ \uparrow, \, \downarrow}}{\gamma}-1}$. In what follows, we develop a perturbation theory treating $\sigma$ as a small parameter. For finite $\sigma$ we look for a solution in the form $\rho_{ \uparrow, \, \downarrow}=\rho_{ 0 \uparrow, \, \downarrow}+q_{ \uparrow, \, \downarrow}$ where $q_{ \uparrow, \, \downarrow}$ is a small perturbation of the real amplitude; $\frac{|q_{ \uparrow, \, \downarrow}|}{\rho_{0 \uparrow, \, \downarrow}} \ll 1$.
Then the equation for $q_{ \uparrow, \, \downarrow}$ is linear and has the form 

 \begin{eqnarray}
    \dot q_{ \uparrow, \, \downarrow}=-\Gamma_{ \uparrow, \, \downarrow} q_{ \uparrow, \, \downarrow} + \frac{\sigma}{\sqrt{1+\bar \alpha^2}} \rho_{0 \downarrow, \, \uparrow} \left( \cos(\theta_{ \downarrow, \, \uparrow} - \theta_{ \uparrow, \, \downarrow} ) -   \alpha  \sin(\theta_{ \downarrow, \, \uparrow} - \theta_{ \uparrow, \, \downarrow} ) \right),
    \label{r_perturb_S}
\end{eqnarray}

where $\Gamma_{ \uparrow, \, \downarrow}=\gamma \frac{2 \rho_{0 \uparrow, \, \downarrow}^2}{1+\rho_{0 \uparrow, \, \downarrow}^2}$. 

Now we assume that the detuning between the unperturbed (free-running) eigenfrequencies of the polarizations is small compared to the effective relaxation rates $\Gamma_{ \uparrow, \, \downarrow}$ of the real amplitudes of the field. Then, the derivative can be neglected in (\ref{r_perturb_S}). Therefore, we assume that the amplitudes follow adiabatically to the variations of the phases of the fields. In this case, the perturbations of the amplitudes can be found as

 \begin{eqnarray}
 q_{\uparrow, \, \downarrow}= \frac{\sigma \rho_{0 \downarrow, \, \uparrow}}{\sqrt{1+ \alpha^2}\Gamma_{ \uparrow, \, \downarrow}} \left( \cos(\theta_{ \downarrow, \, \uparrow} - \theta_{ \uparrow, \, \downarrow} ) - \alpha  \sin(\theta_{ \downarrow, \, \uparrow} - \theta_{ \uparrow, \, \downarrow} ) \right).
    \label{r_perturb_adiab_S}
\end{eqnarray}

Now let us re-write equations (\ref{ampl_phase_eq_S_ampl}) for the phases assuming that $r_{\uparrow, \, \downarrow} \ll \rho_{0\uparrow, \, \downarrow}$

 \begin{eqnarray} 
\dot \theta_{ \uparrow, \, \downarrow}=\omega_{ \uparrow, \, \downarrow} \mp \Omega  +2\left( \mu - \frac{\alpha \gamma }{1+\rho_{0 \uparrow, \, \downarrow}}\right) \rho_{0 \uparrow, \, \downarrow} q_{ \uparrow, \, \downarrow} + 2\left( \tilde \mu - \frac{\tilde \alpha \gamma }{1+\rho_{0 \downarrow, \, \uparrow}}\right) \rho_{0 \downarrow, \, \uparrow} q_{ \downarrow, \, \uparrow}+ \nonumber \\
 + \sigma \frac{\rho_{0 \downarrow, \, \uparrow}}{\sqrt{1+ \alpha^2} \rho_{0 \uparrow, \, \downarrow}} \left( \sin(\theta_{ \downarrow, \, \uparrow} - \theta_{ \uparrow, \, \downarrow} ) + \alpha \cos(\theta_{ \downarrow, \, \uparrow} - \theta_{ \uparrow, \, \downarrow} ) \right). 
 \label{ampl_phase_eq_T_S} 
\end{eqnarray}

where $\omega_{ \uparrow, \, \downarrow}=(\alpha+\tilde \alpha )\gamma+\mu \rho_{0 \uparrow, \, \downarrow}^2 +\tilde \mu \rho_{0 \downarrow, \, \uparrow}^2$ are the free-running frequencies of the polarization states.
Substituting the expression (\ref{r_perturb_adiab_S}) for $q_{ \uparrow, \, \downarrow}$ we obtain equations for the slow dynamics of the phases:

 \begin{eqnarray} 
\dot \theta_{ \uparrow, \, \downarrow}=\omega_{ \uparrow, \, \downarrow} \mp \Omega + \nonumber \\ +\frac{\sigma}{\sqrt{1+ \alpha^2}} \left( \alpha \frac{\rho_{0 \downarrow, \, \uparrow}}{\rho_{0 \uparrow, \, \downarrow}} 
+2\left( \frac{\mu}{\Gamma_{ \uparrow, \, \downarrow} }+ \frac{\tilde \mu}{\Gamma_{ \downarrow, \, \uparrow}} 
-\frac{\alpha \gamma}{\Gamma_{ \uparrow, \, \downarrow} (1+\rho_{0 \uparrow, \, \downarrow}^2 )} 
-\frac{\tilde \alpha \gamma}{\Gamma_{ \downarrow, \, \uparrow} (1+\rho_{0 \downarrow, \, \uparrow}^2 )} \right) \rho_{0 \uparrow, \, \downarrow} \rho_{0 \downarrow, \, \uparrow} \right) \cos(\theta_{ \downarrow, \, \uparrow} - \theta_{ \uparrow, \, \downarrow} ) + \nonumber \\
+\frac{\sigma}{\sqrt{1+ \alpha^2}} \left(   \frac{\rho_{ \downarrow, \, \uparrow}}{\rho_{ \uparrow, \, \downarrow}} 
-2  \alpha \left( \frac{\mu}{\Gamma_{ \uparrow, \, \downarrow} } - \frac{\tilde \mu}{\Gamma_{ \downarrow, \, \uparrow}} 
-\frac{\alpha \gamma}{\Gamma_{ \uparrow, \, \downarrow} (1+\rho_{0 \uparrow, \, \downarrow}^2 )} 
+\frac{\tilde \alpha \gamma}{\Gamma_{ \downarrow, \, \uparrow} (1+\rho_{0 \downarrow, \, \uparrow}^2 )} \right) \rho_{0 \uparrow, \, \downarrow} \rho_{0 \downarrow, \, \uparrow} \right) \sin(\theta_{ \downarrow, \, \uparrow} - \theta_{ \uparrow, \, \downarrow} ).
 \label{phase_eq_1_S} 
\end{eqnarray}

Since we assume that polarizations (different polarization states) are nearly equally excited $\frac{|\rho_{0 \uparrow} - \rho_{0 \downarrow}|}{\rho_{0 \uparrow} + \rho_{0 \downarrow}} \ll 1$  we can disregard their amplitude difference and assume that $\rho_{0 \downarrow} \approx \rho_{0 \uparrow} =\rho_0 $. From this we find that $\Gamma_{ \downarrow} \approx \Gamma_{ \uparrow} = \Gamma$. Then equations for the phases can be reduced to

 \begin{eqnarray} 
\dot \theta_{ \uparrow, \, \downarrow}=\omega_{ \uparrow, \, \downarrow} \mp \Omega +
\frac{\sigma}{\sqrt{1+ \alpha^2}} \left(   \frac{(1+\rho_0^2)(\mu+\tilde \mu)}{\gamma} - \tilde \alpha  \right) \cos(\theta_{ \downarrow, \, \uparrow} - \theta_{ \uparrow, \, \downarrow} )+ \nonumber \\
+\frac{\sigma}{\sqrt{1+ \alpha^2}} \left(  1+ \alpha (\alpha - \tilde \alpha ) - \frac{ \alpha (1+\rho_0^2)(\mu-\tilde \mu)}{\gamma} \right) \sin(\theta_{ \downarrow, \, \uparrow} - \theta_{ \uparrow, \, \downarrow} ).
 \label{phase_eq_2_S} 
\end{eqnarray}

 It is convenient to introduce the mutual phase between the first and the second polarization states as $\varphi=\theta_{ \uparrow}- \theta_{ \downarrow}$. The equation for the phase is then written as
 
 \begin{eqnarray} 
\dot \varphi= \delta - \tilde \sigma \sin \varphi
\label{phase_mut_phase_S} 
\end{eqnarray}
where 
$\delta= \omega_{ \uparrow} - \omega_{ \downarrow}-2\Omega$ is the effective detuning of the free-running frequencies of the polarization states 
and 
\begin{equation}
\tilde \sigma=2\frac{\sigma}{\sqrt{1+ \alpha^2}}\left( 1-\frac{ \alpha (1+\rho_0^2)(\mu-\tilde \mu)}{\gamma} + \alpha (\alpha - \tilde \alpha) \right)
\label{interaction_strength_S}
\end{equation}
is the effective strength of the cross-polarization interaction. The coupling strength (\ref{interaction_strength_S}) can be expressed through the pump $P=P_{\uparrow}\approx P_{\downarrow}$ as
\begin{equation}
\tilde \sigma=2\frac{\sigma}{\sqrt{1+ \alpha^2}}\left( 1-\frac{ \alpha (\mu-\tilde \mu)P}{\gamma^2} + \alpha (\alpha - \tilde \alpha) \right).
\label{interaction_strength_S_through_P}
\end{equation}
One can easily verify that the existence of the fluctuating driving force in (\ref{main_equation_S}) would result in fluctuations of the effective detuning $\delta$.

It is seen that both the nonlinear polariton-polariton interaction and the reservoir nonlinearity contribute to the effective coupling between the polarizations. For some parameter set the $\tilde \sigma$ can become equal to zero which would imply that the second order of the perturbation theory has to be developed. However, it does not correspond to the case discussed in this work.

\section{S2. The fitting of the simulation parameters to the experiment}

To perform numerical simulations we need to fit the parameters $\gamma$, $\alpha$, $\mu$ and $\sigma$  to our experiment. We start with the linear losses in the non-pumped system $\gamma$. Note that we are interested only  in the case where the inter-polarization interaction is weak compared to the dissipation and nonlinear effects. Then, after finding the parameters $\gamma$, $\alpha$, $\mu$ we can disregard the inter-polarization interactions and work with the scalar equation. Consequently, according to our theory, the effective losses depend linearly on the pump intensity $\gamma_{eff}=\gamma(1-\frac{P}{P_{th}})$ if the pump $P$ is below the threshold $P_{th}$. 

\begin{figure*}[t]
    \centering
    \includegraphics[width=0.85\textwidth]{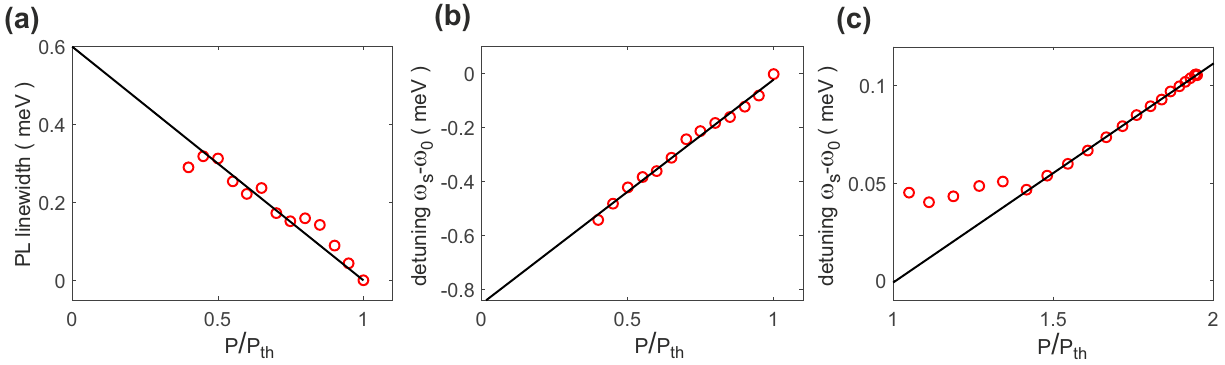}
    \caption{(a) The linewidth of the PL as a function of the pump power (in unit threshold). The data is accurately fitted by a linear fit $\gamma_{eff}=\gamma (1-\frac{P}{P_{th}})$. (b) The PL central frequencies (the energy of the polariton state) for different pump powers. The theoretical fit is $\omega_s=\omega_0+  \eta ( \frac{P}{P_{th}} -1)$ with $\eta=0.83$ where $\omega_0$ is the PL frequency at the threshold. The PL frequencies are shown as their detunings from $\omega_0$.  (c) The experimentally measured frequencies of the condensate above the threshold and the theoretical curve $\omega_s=\omega_0 +\mu ( \frac{P}{P_{th}}-1 )$  for $\mu=0.1125$ meV versus the pump power.  In (b) and (d) the reference frequency $\omega_0$ is the frequency of the condensate at the condensation threshold. In all panels, the open red circles are the experimental data and the black solid lines are the theoretical fits. 
        }
    \label{fig_gamma_bl_shift_Sppl}
\end{figure*}

It is reasonable to assume that the photoluminescence (PL) linewidth is defined by the width of the resonance line which is proportional to the effective losses $\gamma_{eff}$. Assuming that the spectral line is given by a Lorentzian $\frac{1}{(\omega-\omega_s)^2+\gamma_{eff}^2}$, where $\omega_{s}$ - the resonant frequency of the ground state at the given pump, we obtain that the effective losses $\gamma_{eff}$ is a half width of half maximum of the spectral line. The dependence of the experimentally measured PL linewidth on the pump power is shown in Fig.~\ref{fig_gamma_bl_shift_Sppl}(a) by red open circles. The linear fit to these data is shown in the same figure with the black solid line and thus the experimental data is well-reproduced by the mathematical model. The fit allows us to define the losses in the non-pumped system and for our case the losses are $\gamma=0.6$ meV.

Now let us find the parameter $\alpha$ defining the ratio of the blueshift to the gain caused by the reservoir of incoherent excitons. For this we have measured the central frequency of the polariton PL $\omega_{phl}$ as a function of the pump power below the condensation threshold. These results are shown in Fig.~\ref{fig_gamma_bl_shift_Sppl}(b) by red open circles. The theoretical curve of the resonant frequency of the state $\omega_s$ as a function of $P$ is given by $\omega_s=\omega_0+  \eta ( \frac{P}{P_{th}} -1)$ where $\omega_0$ is the eigenfrequency of the state for the threshold pump $P=P_{th}$. The linear fit ($\eta=0.83$ meV) is shown by black solid line in Fig.~\ref{fig_gamma_bl_shift_Sppl}(b), which matches well the experimental data. 

The parameter $\alpha$ is, by definition, the ratio $\alpha=\frac{\eta}{\gamma}\approx \frac{0.83 \rm{meV}}{0.6 \rm{meV} } \approx 1.38$. From the literature it is known that typically $\tilde \alpha \approx -\frac{\alpha}{10}$ and thus in our case $\tilde \alpha \approx - 0.138$.

Above the threshold, the condensation occurs and the intensity of the ground state depends linearly on the pump $|A|^2=\frac{P}{P_{th}}-1$ due to polariton-polariton interaction. It is worth mentioning that above the condensation threshold the reservoir density does not depend on the pump in framework of our model. The experimental dependence of the condensate frequency on the pump above the threshold is shown in Fig.~\ref{fig_gamma_bl_shift_Sppl}(c)  overlapped with the theoretical fit $\omega_s=\omega_0 +\mu ( \frac{P}{P_{th}}-1 )$ for $\mu=0.1125$ meV. The value $\tilde \mu$ can then be estimated as $\tilde \mu=-\frac{\mu}{10}=-0.01125$ meV.

The coupling strength is so small (it lies in the microwave frequency range) that it cannot be measured in optical experiments. To find this value we use our simulation based on the Adler equation (\ref{phase_mut_phase_S}) and find the value of $\tilde \sigma$ given the best agreement with the experiment. With $\gamma$, $\alpha$, $\tilde \alpha$, $\mu$ and $\tilde \mu$ acquired, we can find the value of $\sigma$ using relation (\ref{interaction_strength_S_through_P}). 

We performed numerical simulations of (\ref{main_equation_S}) with the experimental parameters and found an excellent agreement between the results of these full-scale coupled modes equations and the results obtained from the simulations of the corresponding Adler equation.

\section{S3. Viscous motion of a particle}

\begin{figure*}[t]
    \centering
    \includegraphics[width=0.65\textwidth]{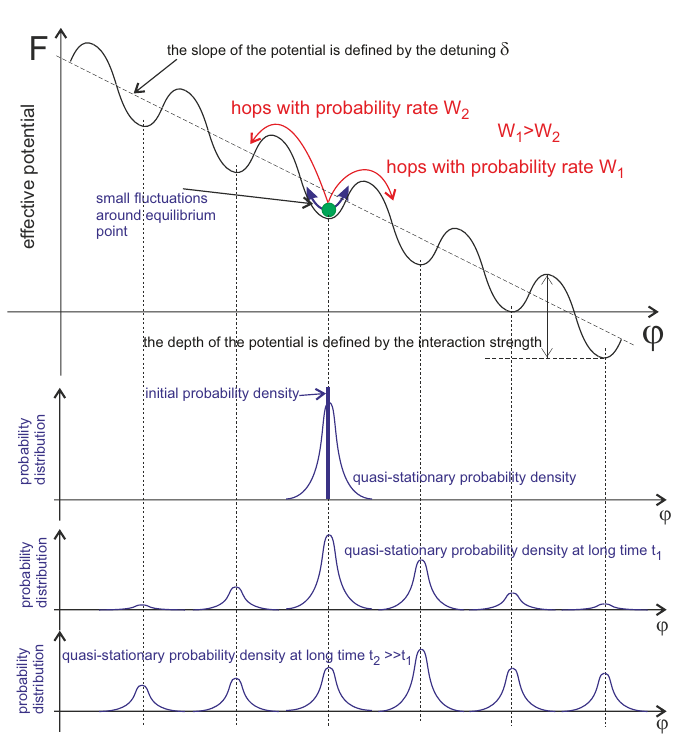}
    \caption{The wash-board potential for equation (\ref{phase_mut_phase_S}). The bottom part of the figure illustrates the temporal evolution of the probability density starting from a delta function (schematically shown by thick blue line. )
    }
    \label{wash_board_Sppl}
\end{figure*}

Now let us qualitatively discuss the evolution of the mutual phase $\varphi$. As mentioned in the main text, this motion is described by the same equation as the viscous motion of a particle in a wash-board potential, see Fig.~\ref{wash_board_Sppl}. Indeed, the equation (\ref{phase_eq_2_S}) can be written as 
 \begin{eqnarray} 
\dot \varphi= -\partial_{\varphi} F(\varphi) 
\label{phase_mut_phase_S_} 
\end{eqnarray}
where $F=-\delta \varphi -\tilde \sigma \cos(\varphi)$ is a wash board potential with the average slope equal to $-\delta$ and the modulation depth equal to $2J$.

There is a critical slope $\delta_{th}=|\tilde \sigma|$ of the washboard potential such that for the small detunings $|\delta|<\delta_{th}$ the potential has maxima and minima. Correspondingly for these detunings, there are stable and unstable equilibrium points,  the relaxation of $\varphi$ to the stable equilibrium point means that the polarization plane rotates at exactly the same velocity as the rotating trap does, the angle between the symmetry axis of the trap and the polarization direction is fixed. At $\delta=\pm \delta_{th}$ the maxima and minima merge and disappear. So for  $|\delta|<\delta_{th}$ the mutual phase infinitely grows in time. In terms of polarization dynamics it means that the polarization plane and the trap rotate with different angular velocities and are not phase-locked.

Let us discuss the evolution of the phase $\varphi$ under the action of weak random variations of the potential slope (fluctuations of $\delta$ with zero mean value) for $|<\delta>|<\delta_{th}$. Then the phase $\varphi$ is characterized by the probability density function.  If the noise intensity goes to zero the probability density becomes a delta-function situated at $\varphi=\text{arcsin}( \frac{\delta}{\tilde \sigma} )$.

Switching the noise on leads to the expansion of the probability density, the width of the probability density function is defined by the intensity of the noise and the curvature of the potential. The evolution of the probability density for the viscous motion in a washboard potential is illustrated in Fig.~\ref{wash_board_Sppl}.
It is known that, because of the broadening of the probability density function, the correlation function of $\exp(i \varphi)$ decreases. 

If potential $F$ has such a deep minima so the "hops" ( see Fig.~\ref{wash_board_Sppl} ) from one minima to another are negligible then the probability density has a localized stationary distribution, correspondingly the correlation function decreases to a non-zero constant at characteristic time equal to the "broadening time" of the probability density. It is worth noting that the absence of the "hops" is actually equivalent to approximation of the washboard potential by an infinitely deep potential. Then the broadening of the probability density function is limited and the noise does not destroy "far ordering" in terms of solid-state physics (crystallinity survives). Thus, in our case, this would correspond to the formation of time crystals. The analogy is Laue diffraction of X-rays (or electrons, neutrons) on a crystal: the temperature does not result in the broadening of the spectral lines of the scattered field, but around each of the line a bell-shaped background appears [27]. In radiophysics, this is known as the phase noise.    

In a real potential the probability of the "hops" from one potential minima to another is finite. The hop to the right (referring to the scheme in Fig.~\ref{wash_board_Sppl}) is more probable than the hop to the left and thus the average $\varphi$ is growing in time. This means that, because of the action of the noise, the average pseudo-spin precession frequency is not equal exactly to $2\Omega$ and experience fluctuations. The hops destroy the far ordering (it can also be referred as frequency noise). Consequently, the correlation function finally decays to zero. 

However, for relatively deep minima of the potential it is possible to distinguish two different characteristic times of the correlation function evolution. The first one is the characteristic probability density spreading time within one minimum of the potential. After this time the correlation function drops from its maximum value to a slowly decaying plateau. For our experiments, this time is of the order of several nanoseconds. The second time is defined by the rates of the hops to the neighbouring potential minima and the corresponding further spreading of the probability distribution at the distances exceeding the distance between the potential minima. This process results in the slow decay of the correlation function plateau to zero. For our experiments, the second characteristic time (hundreds of nanoseconds) is many orders of magnitude longer than the first characteristic time.

\section{S4. Resonance curves for different coupling strength}

\begin{figure*}[t]
    \centering
    \includegraphics[width=0.45\textwidth]{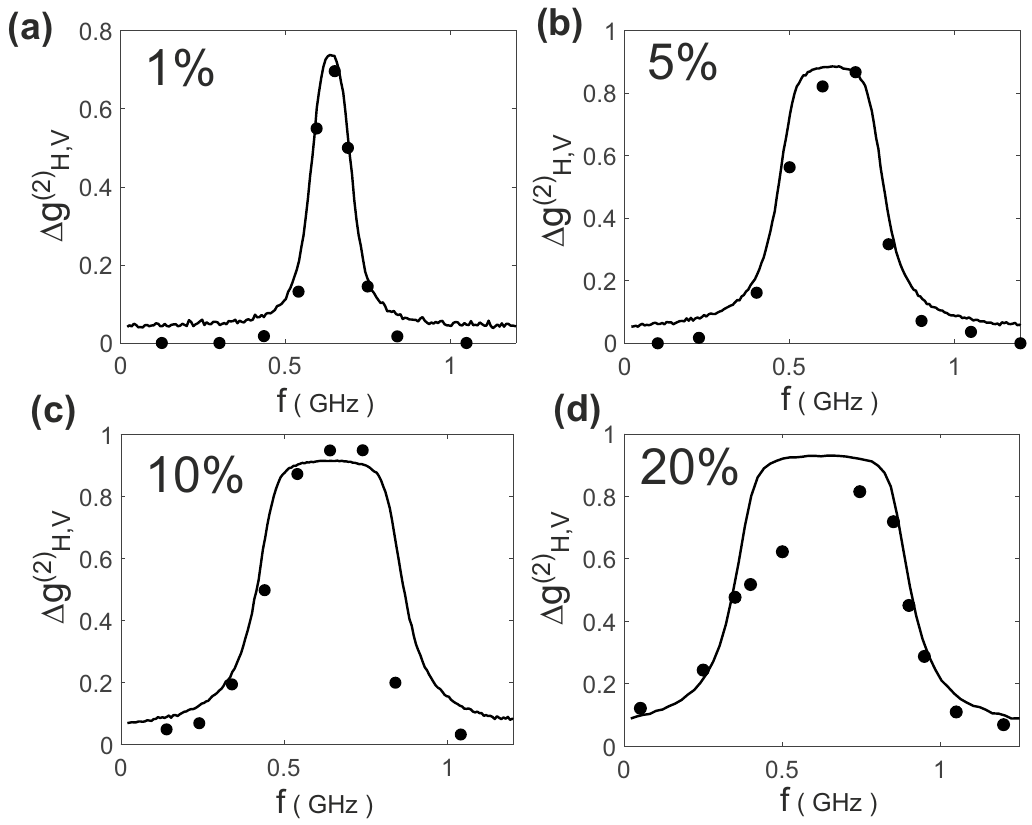}
    \caption{$\Delta g^{(2)}_{H,V}$ as a function of the stirring frequency $f$ measured and calculated for different $r$. Panels (a)-(d) show the experimental and simulated data for the ratios $r=1\%, 5\% ,10\%$ and $20\%$ with corresponding $\tilde \sigma=0.6$ $\mu$eV ( $144$ MHZ), $\tilde \sigma=1.28$ $\mu$eV ( $310$ MHz ), $\tilde \sigma=1.73$ $\mu$eV ( $418$  MHz) and $\tilde \sigma=2.24$ $\mu$eV ( $542$ MHz ) respectively. The experimental data are shown with black dots, the simulation results - with solid black curves. 
    }
    \label{res_curves_Sppl_}
\end{figure*}

In order to theretically find the cross-polarization interaction strength accounted by parameter $\sigma$ we fit the experimental measurements of radio-frequency dynamics of the condensate polarization. We tune the coupling strength to get the best fit of the synchronization range for $r=20$\% because in this case, the effect of noise is less compared to the weaker coupling strengths, see Fig.~\ref{res_curves_Sppl_}(d). The coupling strength $\tilde \sigma=2.24$ $\mu$eV ($542$ MHz) provides a good fit. 

To find the coupling strengths for other ratios $r$ we use the following reasoning. The coupling strength is proportional to the product of the TE-TM splitting and the depth of the angular modulation of the reservoir. One can expect that TE-TM splitting does not depend on $r$, as long as the shape of the condensate is not strongly affected. We estimate the modulation depth of the reservoir provided that the average density of the reservoir is fixed. The intensity of the potential is given by $| \sqrt{I_1}e^{i\Theta-i\Omega t} +\sqrt{I_2}e^{-i\Theta+i\Omega t]}   |^2W(\zeta)$ where $I_1$ and $I_2$ are the intensities of the pump beam components with angular indexes $l=\pm 1$, $\Theta$ is the angular coordinate, $W(\zeta)$ describes the beam intensities distributions along radial coordinate $\zeta$. 
The angular independent and angular dependent parts (with zero average) of the pump can then be written as $F_p=I_1+I_2$ and $F_o=2\sqrt{I_1}\sqrt{I_2} W \cos(2(\Theta-\Omega t))$ correspondingly. Knowing the ratio of the intensities one can express the intensity of the second beam as $I_2=rI_1$. We obtain that $F_p=(1+r)I_1W$ and $F_o=2I_1\sqrt{r}W\cos(2\Theta)$. Keeping the average pump constant we find that the amplitude of the angular dependent pump depends on $r$ as $F_o=2F_p \cos(2\Theta) \frac{\sqrt{r}}{1+r}$. 

Using this ratio we can find the coupling strength for all our experimental values of $r$ knowing $\tilde \sigma$ for $r=20$\%. Thus we find that the coupling strength should be $\tilde \sigma=0.6$ $\mu$eV for $r=1$\%, $\tilde \sigma=1.28$ $\mu$eV for $r=5$\% and $\tilde \sigma=1.73$ $\mu$eV for $r=10$\%. We also tune the intensity of the noise to get the best fit to the shape of the curve for $r=1$\% because in this case the effect of noise is maximal. The theoretical and experimental dependencies of $\Delta g^{(2)}_{H,V}$ on $f$ are shown in Fig.~\ref{res_curves_Sppl_} for different ratios $r$. The theory fits the experiment well.

\section{S6. The effect of the noise on the spectrum of the intensity of optical radiation  measured through a linear polarizer }

Let us discuss here how the developed theory reducing the problem to a simple Adler equation allows us to describe the spectrum of the radiation measured through a linear polarizer. As it is discussed in the main text the intensity of such radiation can be expressed as  
$I=2\rho_0^2 (1+\cos(2\Omega t +\varphi))$. We will look for the oscillating part of the signal $I_{osc}=2\rho_0^2 \cos( 2\Omega t +\varphi) $. We neglect fluctuation of the amplitude which is justified for pumps of high intensity, but account for the fluctuations of the phase $\varphi$. The goal of the present section is to find the spectrum of $I$ for $\Omega$ lying within the synchronization regime.

The equation for $\varphi$ is 
 \begin{eqnarray} 
\dot \varphi+ \tilde \sigma \sin \varphi=\delta_{0} +\delta_f,
\label{phase_mut_phase_fluct} 
\end{eqnarray}
where $\delta_0$ is the average difference of the polarizations free-running frequencies and $\delta_f$ if the fluctuations of the frequencies difference. We assume $\delta_f$ to be $\delta$-correlated stationary noise with the spectrum $S(\omega)=S_0$.

To proceed we represent the mutual phase as $\varphi=\varphi_0 +\varphi_f$ where $\varphi_0$ is a stationary solution without noise (so $\tilde \sigma \sin\varphi_0 =\delta_0$). Then we assume that the fluctuations are weak so that the deviation of $\varphi$ from $\varphi_0$ can be considered to be small and the hopes to the neighbouring equilibrium points are negligible rare events. Then a linear equation can be written for $\varphi_f$
 \begin{eqnarray} 
\dot \varphi_f+ \Delta_S \varphi=\delta_f.
\label{phase_mut_phase_fluct_lin} 
\end{eqnarray}
where $\Delta_S=\tilde \sigma \cos(\varphi_0)$. From here it is straightforward to write the expression for the spectrum of $\varphi_f$
$$S_{\varphi}=\frac{S_0}{\omega^2 +\Delta_S^2}$$

For small $\varphi_f$ the measured oscillating signal can be represented as $$I_{osc}=2\rho_0^2 ( \cos( 2\Omega t + \varphi_0 )- \varphi_f \sin( 2\Omega t +\varphi_0) ) $$ The spectra of the first term is a $\delta$-function at $\omega=2\Omega$. Now we analyze the spectrum of the second component, and denote it as $I_{n}$ $$I_n=2\rho_0^2  \varphi_f \sin( 2\Omega t +\varphi_0).$$ 
The spectrum of $I_n$ can be expressed as $S_{In}= \rho_0^4\left(S_{\varphi}(\omega-2\Omega) +S_{\varphi}(\omega+2\Omega)\right)$.

So we can conclude that in the regime of synchronization, the spectrum should consist of a $\delta$-function situated at $2\Omega$ and a pedestal produced by the fluctuations of the phase. The shape of the pedestal is given by
 \begin{eqnarray} 
S_{In}=\rho_0^4 S_0 \left( \frac{1}{(\omega-2\Omega)^2 + \Delta_S^2 }+\frac{1}{(\omega+2\Omega)^2 + \Delta_S^2 } \right).
\label{pedesta_spectrum} 
\end{eqnarray}
So we can expect that the pedestal in the spectrum is a Lorenz-shaped spectral line. It is interesting to note that the increase of the synchronization strength $\tilde \sigma$ results is the broadening of the pedestal and the decrease of its height. The width of the pedestal is maximal and the height is minimal at the center of the synchronization domain $2\Omega=\delta_0$ where $\Delta_S$ reaches its maximum equal to $\tilde \sigma$. Let us also remark that the developed perturbation does not work at the edges of the synchronization range where the fluctuations of the phase are significant. 

\begin{figure*}[t]
    \centering
    \includegraphics[width=0.45\textwidth]{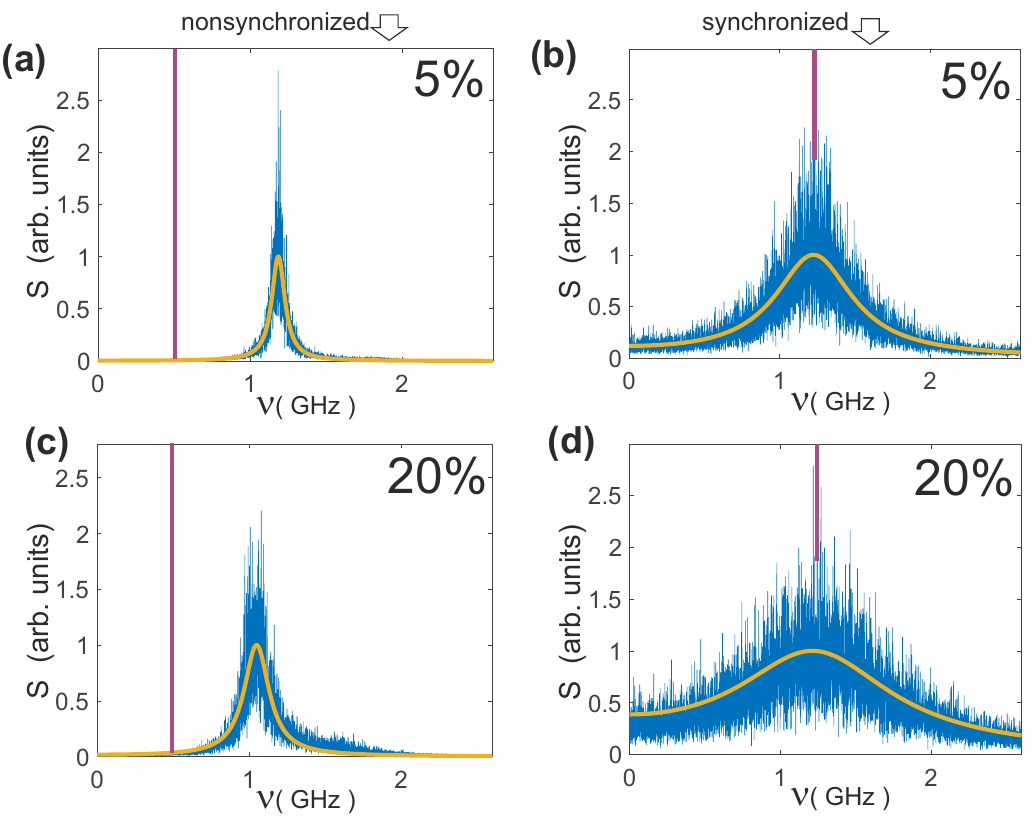}
    \caption{ Numerically calculated spectra averaged over $7$ realizations are shown in (a,b) for $r=5\%$ and in (c,d) for $r=20\%$ for different rotation frequencies of the pump $f$. The straight purple lines are the $\delta$-function-like components of the spectra (our calculations do not allow us to resolve this line). The positions of these purple lines can also be seen as a mark of the frequency equal to $2f$ (double angular velocity of the pump). In all panels the yellow curves show the Lorentzian fits to the pedestals of the numerically calculated spectra. The spectra are normalized such that the maximum of the Lorentzian fit is $1$. 
    }
    \label{spectra_noise_Sppl}
\end{figure*}

Now let us compare the results of numerical simulations against the analytical results. The numerically calculated spectra averaged over $7$ realizations are shown in Fig.~\ref{spectra_noise_Sppl}. Then we filtered out the narrow $\delta$-function line from the numerically calculated spectrum and fitted the residual pedestal with the Lorentzian spectral line. It is seen that the Lorenzian curves are good approximations for the spectra observed in numerical simulations.  Let us remark that out of the synchronization, the spectrum still contains a narrow (in ideal case  $\delta$-function ) line appearing because of the periodically rotating potential, see Figs.~\ref{spectra_noise_Sppl}(a,c), however, the intensity of this line is much less compared to that in the synchronization regime.  

\begin{figure*}[t]
    \centering
    \includegraphics[width=0.45\textwidth]{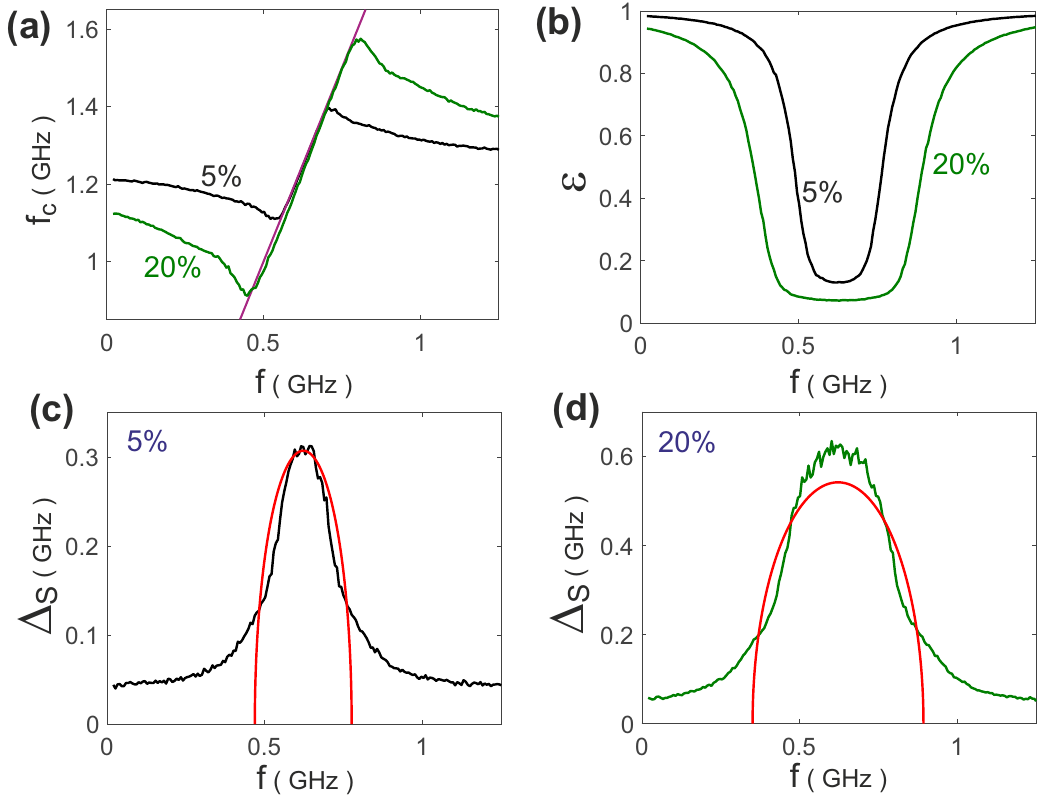}
    \caption{ (a) The position of the spectra pedestal maxima (obtained by fitting the pedestals of the numerically calculated spectra by the Lorentzian lines) as a function of the rotation velocities of the trap for $r=5\%$ (black line) and $r=20\%$ (green line). The magenta straight line $f_c=2 f$ shows the expected position of the spectral maximum in the regime of synchronization. Panel (b) shows the numerically found ratio $\varepsilon$ of the energy contained in the pedestal to the total energy of the spectrum. The width of the pedestals as functions of the trap rotation velocity $f$ are shown in (c), (d) for $r=5\%$ and $r=20\%$, respectively. The black and green curves are the results of numerical simulations. The analytical dependencies  of the spectrum width $\Delta_S$ on the trap rotation velocity $f$ found within the synchronization range are shown by the red curves in (c) and (d).  
    }
    \label{spectra_param_Sppl}
\end{figure*}

To shed light on the synchronization we traced how the position of the spectrum pedestal maximum depends on the rotation velocity of the trap $f$. To do this we use the fitting of the numerically calculated pedestal by the Lorenzian spectral line and use the fit to find the central frequency $f_{c}$. The dependencies are shown in  Fig.~\ref{spectra_param_Sppl}(a) for $r=5\%$ and $r=20\%$. It is evident that for $f$ in the synchronization range the position of the pedestal maximum follows the rotation velocity of the trap.

The numerically found ratio $\varepsilon$ of the energy in the spectrum pedestal (with the filtered out $\delta$-function - like spectral line) to the total energy in the spectrum is shown in Fig.~\ref{spectra_param_Sppl}(b). This confirms that the synchronization leads to the formation of intense very narrow spectral line. This is equivalent to the formation of a relatively high and very slowly decaying plateau on the correlation functions ($g^{(2)}_{H,V}$) in the main text.

Finally, we investigate how the width of the spectrum defined as  $\Delta_S$ of the Lorentzian fit depends on the rotation velocity $f$, see Fig.~\ref{spectra_param_Sppl}(c,d) showing these dependencies for $r=5\%$ and $r=20\%$. It is seen that the numerically calculated spectra have maxima at the center of the synchronization ranges. Note, that for a stronger coupling the pedestal of the spectra is wider. The analytical dependencies of the spectra width on $f$ for the experimental parameters are shown in Fig.~\ref{spectra_param_Sppl}(c,d) with the red lines. Close to the center of synchronization range the analytics gives a reasonably good fit to the numerics.

\section{S7. Frequency response of the HBT setup.}

\begin{figure}[h]
    \centering
    \includegraphics[width=0.6\columnwidth]{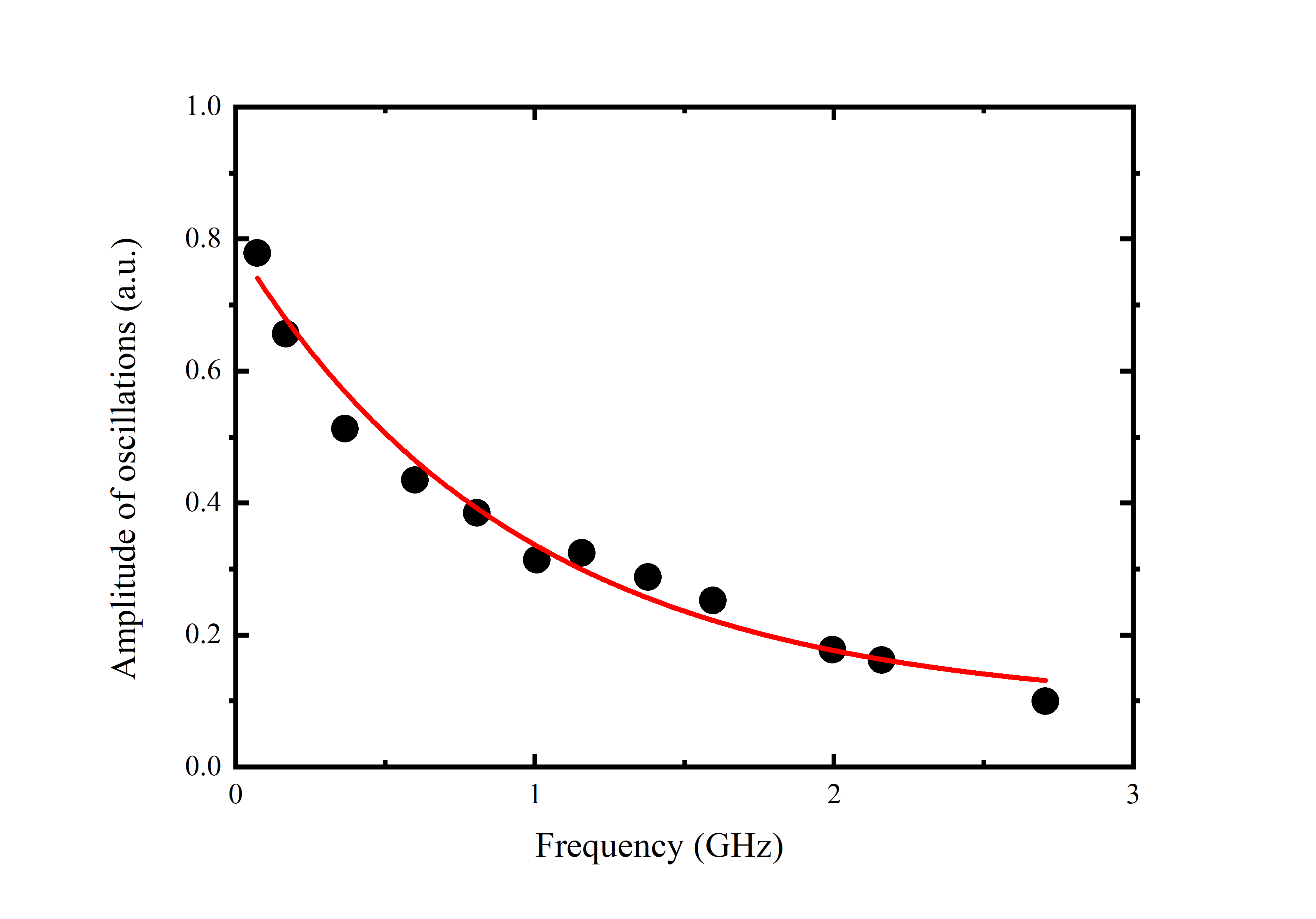}
   \caption{The frequency response of the HBT setup. The amplitude of the intensity auto-correlation function $g^{(2)}_{Total}$ for the two beating lasers versus their frequency difference (beating frequency). The red line is an exponential fit of the experimental data.
   }
    \label{response}
\end{figure}%

In the experiment, we observe the decrease in the amplitude of the $g^{(2)}_{H,V}$ oscillations at higher GHz rotations of the optical trap (see Figs. 3(a,b) in the main text). In order to quantitatively characterize this effect, we measure the frequency response of the utilized detection system (HBT interferometer). We make the intensities of the two excitation lasers equal, and vary their frequency detuning while measuring the $g^{(2)}$ of the obtained beating note. As a result, we obtain the dependence of the amplitude of the $g^{(2)}$ oscillations on the frequency detuning, as depicted in Fig.~\ref{response}. The observed decrease of the oscillation amplitude is in agreement with the measurement of the  $g^{(2)}_{H,V}$ of the condensate emission at GHz stirring. We attribute this observation to the technical limitations of our detection system (the frequency bandwidth of the TSCPC cards, characteristics of single photon counting avalanche photodiodes, etc.). Furthermore, we utilize the curve shown in Fig.~\ref{response} in order to compensate for the frequency response of our setup. We renormalize the experimentally obtained ranges of $g^{(2)}_{H,V}$ with respect to the measured frequency of the spin oscillations. As a result, we obtain the resonance curves on the stirring frequency depicted in Figs.~1(c) and 1(d) in the main text.

\section{S8. Precession resonance for r= 30 percent}

\begin{figure}[h]
    \centering
    \includegraphics[width=0.5\columnwidth]{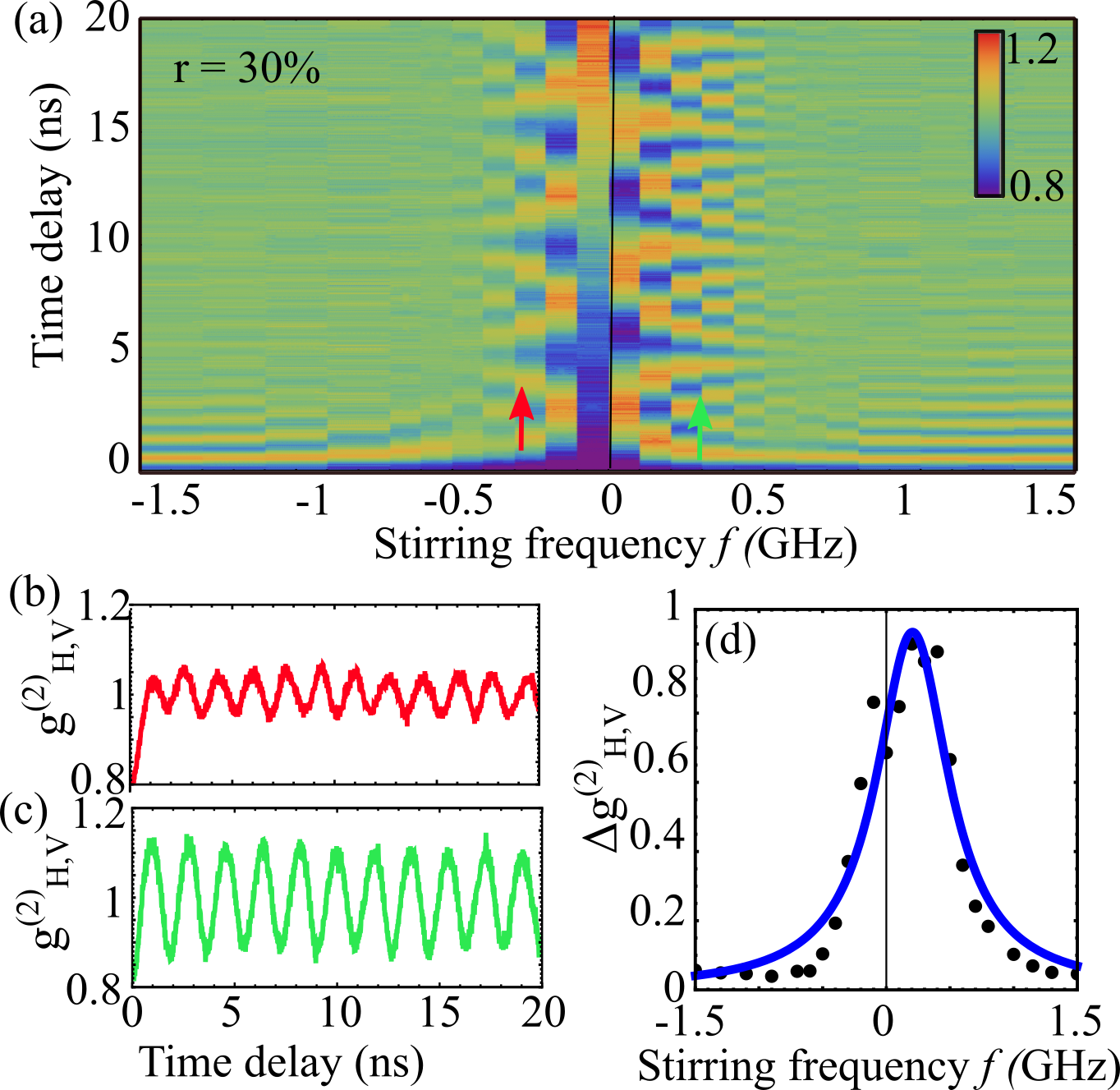}
    \caption{(a) $g^{(2)}_{H,V}$ as a function of the stirring frequency $f$ and time delay of the HBT interferometer for $r= 30\%$. The inset in panel (a) depicts the range of the $g^{(2)}_{H,V}$ in the 2 ns vicinity of 15 ns time delay. (b) and (c) are cross-sections of panel (a) at $f = -0.3$ GHz and $f = 0.3 $ GHz, schematically marked by red and green arrows. (d) Measured corresponding resonant dependence of the $\Delta g^{(2)}_{H,V}$ on the stirring frequency $f$ (black dots) fitted by a Lorentzian profile  (blue curve). Note the wider frequency range compared to Fig.~S3.}
    \label{fig4}
\end{figure}

For the case of the maximal studied $r = 30\%$ we observe the resonance close to the zero stirring frequency, as visible from Fig.~\ref{fig4}(a). The modulation of the rotating trap is so big that it affects the state of the condensate, widening the range of the stirring frequencies resulting into the resonance response of the precession. Moreover, the resonance spans for both clock and counterclockwise stirring directions (see Figs.~\ref{fig4}(b-d)). The driven precession is completely independent of the direction of the self-induced spin precession. This result cannot be reproduced by the numerical simulations since our perturbation approach does not take into account the poor confinement and losses in the condensate for the bigger pump ratio $r$. However, we note that the width of the resonance is still well described by the developed theory.

\section{S9. Mutual phases measurements}

\begin{figure}[h]
    \centering
    \includegraphics[width=0.5\columnwidth]{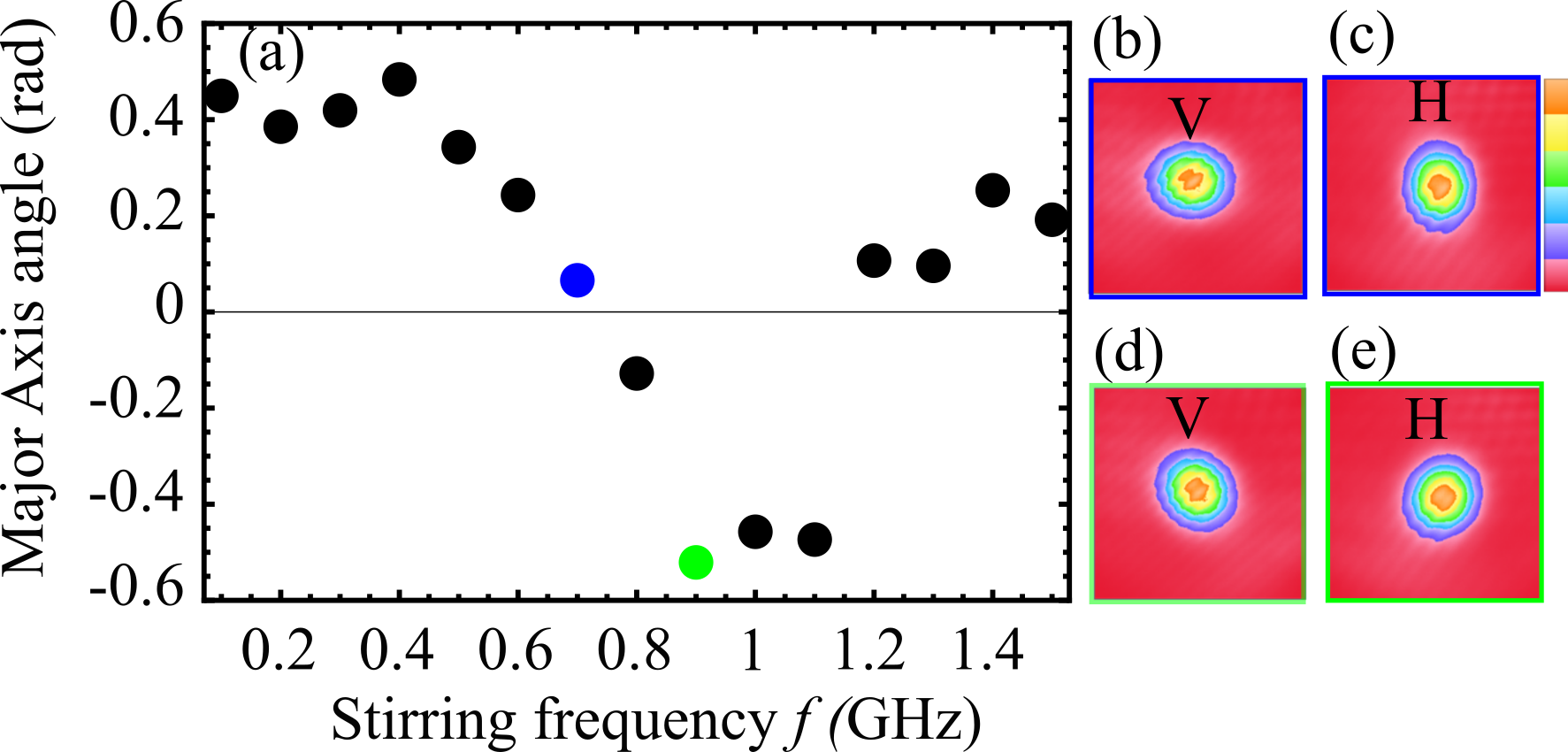}
    \caption{ (a) The major axis angle for the V condensate projection at the different stirring frequencies. (b,d) are the vertical (V) polarization projections of the condensate density at $f = 0.7$ GHz and $f = 0.9$ GHz, respectively. (d,e) are the horizontal (H) polarization projections of the condensate density at $f = 0.7$ GHz and $f = 0.9$ GHz, respectively. The blue and the green points in panel (a) correspond to the major axis angle, retrieved for the condensate polarization projections depicted in panels (b) and (d), respectively.}
    \label{angle}
\end{figure}

To experimentally measure the mutual phase between the polarization components we split the condensate PL with the polarizing beamsplitter and project it on camera. As a result, we obtain the projections of the split modes on the horizontal (H) - vertical (V) basis. The change of the mutual phase of the split components with respect to the stirring frequency (depicted in Fig.~3 in the main text) results in differently aligned projections of the condensate on the H-V basis. For example, at the precession resonance (here $r = 20\%$ and $f = 0.7$ GHz), the mutual phase is zero and we observe the horizontally and vertically elongated elliptical condensate in V and H projections, respectively (see Figs.~\ref{angle}(b) and \ref{angle}(c)). Note, that as discussed in the main text and in Ref.~[2] the elliptical shape of the condensate comes from the ellipticity of the confining and rotating potential. The non-zero mutual phase at a higher stirring frequency further from the resonance results in the tilt of the elliptical condensate projections with the major axis angle corresponding exactly to the mutual phase between the polarization components. In Figs.~\ref{angle}(d) and \ref{angle}(e) we plot the V, H condensate projections at $f = 0.9$ GHz, one can observe that they are slightly inclined. We retrieve the major axis angle for the V projection by fitting the condensate density with 2D Gaussian and plot the retrieved angle for the different stirring frequencies in Fig.~\ref{angle}(a). The clear dependence in the synchronization range for $f$ from 0.4 to 0.9 GHz is observed. We emphasize, that outside the synchronization range, both projections are almost cylindrically symmetric making the retrieval of the major axis angle challenging and not precise, that is why in the main text we plot only the phases in the synchronization range. 

\end{document}